\documentclass[aps,prb,preprint,showpacs,superscriptaddress]{revtex4}
\usepackage{graphicx}
\begin{document}
\title{Structural determination of a low-symmetry surface by low-energy electron diffraction and {\it ab initio} calculations: Bi(110)}
\author{J. Sun}
\email[Electronic address: ]{jsun@cisunix.unh.edu}
\affiliation{Department of Physics and Materials Science Program, University of New Hampshire, Durham, New Hampshire 03824, USA}
\author{A. Mikkelsen}
\altaffiliation[]{Permanent address: Department of Synchrotron Radiation, Lund University, SE-22100 Lund, Sweden}
\affiliation{Institute for Storage Ring Facilities and Interdisciplinary Nanoscience Center (iNano), 
University of Aarhus, 8000 Aarhus C, Denmark}
\author{M. Fuglsang Jensen}
\affiliation{Institute for Storage Ring Facilities and Interdisciplinary Nanoscience Center (iNano), 
University of Aarhus, 8000 Aarhus C, Denmark}
\author{Y.M. Koroteev}
\affiliation{Donostia International Physics Center (DIPC),
20018 San Sebasti{\'a}n, Basque Country, Spain}
\affiliation{Institute of Strength Physics and Materials  Science, Russian
Academy of Sciences, 634021, Tomsk, Russia}
\author{G. Bihlmayer}
\affiliation{Institut f\"ur Festk\"orperforschung, Forschungszentrum
J\"ulich, D-52425 J\"ulich, Germany}
\author{E.V. Chulkov}
\affiliation{Donostia International Physics Center (DIPC),
20018 San Sebasti{\'a}n, Basque Country, Spain}
\affiliation{Departamento de F\'{\i}sica de Materiales and Centro Mixto CSIC-UPV/EHU,
Facultad de Ciencias Qu\'{\i}micas, UPV/EHU, Apdo.~1072, 20080
San Sebasti{\'a}n, Basque Country, Spain}
\author{D.L. Adams}
\affiliation{Institute for Physcis and Astronomy, University of Aarhus, 8000 Aarhus C, Denmark}
\author{Ph. Hofmann}
\affiliation{Institute for Storage Ring Facilities and Interdisciplinary Nanoscience Center (iNano), 
University of Aarhus, 8000 Aarhus C, Denmark}
\author{K. Pohl}
\affiliation{Department of Physics and Materials Science Program, University of New Hampshire, Durham, New Hampshire 03824, USA}
\date{\today}

\begin{abstract}
The surface structure of Bi(110) has been investigated 
by low-energy electron diffraction (LEED) intensity analysis and by first-principles calculations.  
Diffraction patterns at a sample temperature of 110 K and normal incidence 
reveal a bulk truncated (1$\times$1) surface without indication of any structural reconstruction 
despite the presence of dangling bonds on the surface layer. 
Good agreement is obtained between the calculated and measured diffraction intensities 
for this complex, low-symmetry surface containing only one mirror-plane symmetry element. 
No significant interlayer spacing relaxations are found. 
The Debye temperature for the surface layer is found to be lower than in the bulk, 
which is indicative of larger vibrational atomic amplitudes at the surface. 
Meanwhile, the second layer shows a Debye temperature close to the bulk value. 
The experimental results for the relaxations agree well with those of our first-principles calculation.
\end{abstract}


\maketitle

\section{INTRODUCTION}
Characteristic of group V elements, bismuth crystallizes in the rhombohedral A7 structure as a semimetal 
with a small density of states at the Fermi level.\cite{Issi:1979}
But interestingly, the surfaces of Bi show very different electronic properties than the bulk. 
Studies on the Bi(110)\cite{Agergaard:2001}, Bi(100)\cite{Hofmann:2005} and
Bi(111)\cite{Hengsberger:2000,Ast:2001} surfaces have shown that they are much more metallic than the bulk 
due to a significantly higher density of states at the Fermi level at the surface.\cite{Hofmann:2006}
It has been found that one significant contribution is from a strong spin-orbit coupling at the surface 
due to broken inversion symmetry.\cite{Agergaard:2001,Koroteev:2004,Pascual:2004,Hofmann:2006}

From a chemical point of view, the creation of a surface requires the breaking of atomic bonds. 
Covalent bonding plays only a minor role in most metals. 
Thus the effect of bond-breaking is small and surface properties are similar to those of the bulk, 
although localized electronic surface states may be present. 
On semiconductors, creating a surface leaves so-called dangling bonds which should give rise to
half-filled and therefore metallic bands. 
However, it turns out that on most semiconductor surfaces the atoms re-arrange their positions 
such that the dangling bonds are removed and the surface is again a semiconductor and not a metal.\cite{Jona:1979}
Semimetals such as bismuth lie in between these two cases. 
On one hand, a semimetal is close to being a semiconductor 
since directional bonding is important and the valence and conduction bands are almost separated by a gap. 
On the other hand, there is a very small overlap between both bands 
such that the material is formally a metal.
This delicate balance between being a metal and a semiconductor depends crucially on the atomic structure\cite{Shick:1999} 
and it can be expected to be severely disturbed at the surface. 

Detailed structural information on Bi surfaces is so far limited to a recent LEED intensity-vs-voltage (IV) 
and first-principles study of the Bi(111) surface.\cite{Monig:2005} 
One important difference between bulk terminated Bi(110) and Bi(111) is that 
the Bi(110) surface exhibits dangling bonds, while Bi(111) does not.
In a pioneering study by Jona,\cite{Jona:1967}
oxygen adsorption experiments suggest that Bi(110) is noticeably more active than Bi(111).
A qualitative analysis of LEED patterns in Jona's study shows an unreconstructed (1$\times$1) Bi(110) surface structure.
From the bulk structure Jona erroneously concluded that the unit cell (and hence the LEED pattern) 
should not be exactly rectangular but that the lattice vectors should include an angle 
slightly different from 90$^{\circ}$.
This is not correct, as will become appartent  below. 
The unit cell is rectangular and almost quadratic.  
A recent scanning tunneling microscopy study by Pascual \textit{et al.},\cite{Pascual:2004}
revealed images of the Bi(110) surface that are consistent with a near-square surface unit cell.

A surface structural analysis of this surface has not been undertaken before because of its very low symmetry.
The Bi(110) surface has only one symmetry element, a mirror plane, which makes the LEED-IV analysis challenging.
An additional complication for Bi(110) is the close stacking of atomic layers, 
which requires an initial calculation of the bulk diffraction matrices via the combined space method.\cite{VanHove:1979}
To our best knowledge, this situation has not been encountered in LEED-IV analyses. 

In this article, we report on a study of the surface structure of clean Bi(110) 
by quantitative LEED intensity-vs-voltage analysis and {\it ab initio} calculations.
Experimental  diffraction intensities taken at a sample temperature of 110 K under normal incidence 
have been analyzed by comparison to dynamical LEED calculations.
Great care was taken to align the sample considering the low symmetry diffraction pattern.
The main structural parameters that were optimized in the LEED-IV analysis 
include the first 4 interlayer spacings and the Debye temperatures for the first 2 surface layers.
Furthermore, we have performed first-principles calculations for the atomic structure of Bi(110). 
The results are in good agreement with the experimental relaxations.
The Bi(110) surface can serve as an example for the application of LEED intensity analysis to low symmetry systems. 

In the following, we introduce the bulk truncated surface structure of Bi(110). 
Experimental and computational details are described in Sec.~\ref{sec:3333}, 
followed by the results and discussion in Sec.~\ref{sec:5555}. 
Conclusions are given in Sec.~\ref{sec:6666}.

\section{BULK TRUNCATED BI(110) SURFACE STRUCTURE} \label{sec:2222}

\begin{figure}
\includegraphics[width=3.25in]{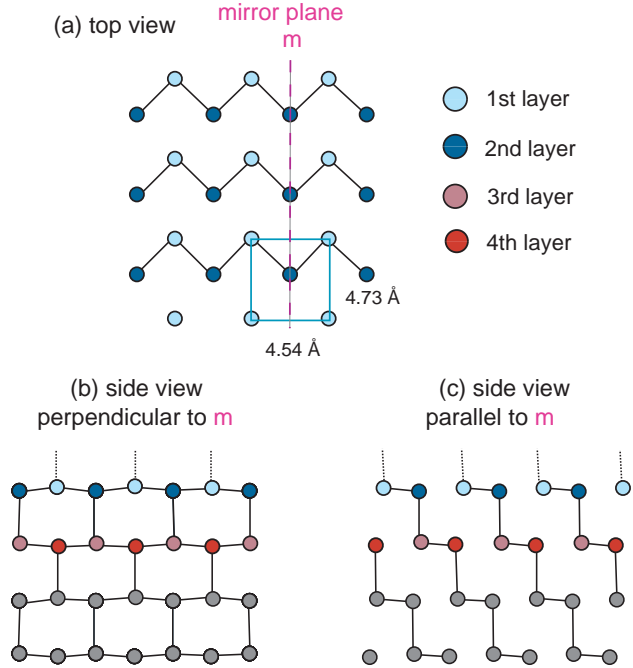}
\caption{(color online) Truncated-bulk structure of Bi(110). 
The solid lines and dotted lines mark covalent bonds and dangling bonds, respectively. 
(a) Top view of the first two atomic layers.
Each layer consists of a two-dimensional rectangular lattice and the lattice constants at 110 K are given. 
The mirror planes of the structure are also shown as dashed lines. 
(b) and (c) Side views of the first eight layers perpendicular and parallel to the mirror plane, respectively. 
The bilayer-like structure with alternating short and long interlayer spacings is evident.}\label{fig:stru}
\end{figure}

The A7 ($\alpha$Arsenic) structure of bulk bismuth has a rhombohedral unit cell with a two-atom basis.
The bulk truncated surface structure of Bi(110) is shown in Fig.~\ref{fig:stru}. 
The side views show the bilayer stacking
with a single bond connecting every other atom between neighboring bilayers. 
Within one bilayer, each atom in one layer bonds with two nearest-neighbor atoms in the other layer. 
The covalent bonds have been drawn by solid lines and the dangling bonds at the surface layer by dashed lines. 
The bilayer-type structure gives rise to alternating interlayer distances. 
For the truncated bulk at 110 K we have:
$d_{12}^{\mathrm{b}} = 0.208$ \AA, 
$d_{23}^{\mathrm{b}} = 3.064$ \AA, 
$d_{34}^{\mathrm{b}} = 0.208$ \AA, 
$d_{45}^{\mathrm{b}} = 3.064$ \AA, 
and so on.
Interlayer spacings between the $i$th and $j$th bulk layers are indicated as $d_{ij}^{\mathrm{b}}$. 
Noticeably, the Bi(110) surface has very low symmetry: 
the only symmetry element is a mirror plane as indicated in Fig.~\ref{fig:stru}. 
The lengths of unit vectors at 110 K are taken as 4.731 \AA\ and 4.538 \AA; 
see Refs.~\onlinecite{Cucka:1962, Jona:1967, Liu:1995}.
If the rhombohedral structure is treated as a pseudocubic structure as in Ref.~\onlinecite{Jona:1967}, 
Bi(110) will be denoted as Bi(100). 
The pseudo-square character of the surface unit cell is evident: 
for a cubic Bi structure all the atoms in the first bilayer would have the same height, 
the unit cell would be rotated by about 45$^{\circ}$, and contains only one atom. 

\section{METHODS}\label{sec:3333}
\subsection{Experimental}

The experiment was performed in  a $\mu$-metal ultrahigh vacuum chamber equipped with a four-grid LEED optic
with a base pressure of 7$\times$10$^{-9}$ Pa.  
Surface contamination was measured by Auger Electron Spectroscopy (AES) using a hemispherical electron analyzer 
and the LEED electron gun as electron source. 
The sample was mounted on a manipulator, allowing positioning to within $0.1^{\circ}$ around all three axes of the crystal. 
The sample was cooled by liquid nitrogen. 
The surface was cleaned by cycles of 1 keV Ar$^{+}$ sputtering and annealing to $150^{\circ}$C.
With AES no surface contamination could be detected. 
The maximum possible oxygen contamination was determined to be 0.02 monolayers.
Spot intensities were measured using a 16 bit Charge-Coupled Device (CCD) camera. 
A back-illuminated and Peltier cooled ($-40^{\circ}$C) CCD chip guaranteed a high quantum efficiency
and low dark current. 
The camera was mounted on a base, which allowed rotation around all three axes. 
Great care was taken to align the camera with respect to the electron gun and the Bi crystal, as described below.

\begin{figure}
\includegraphics[width=5.0in]{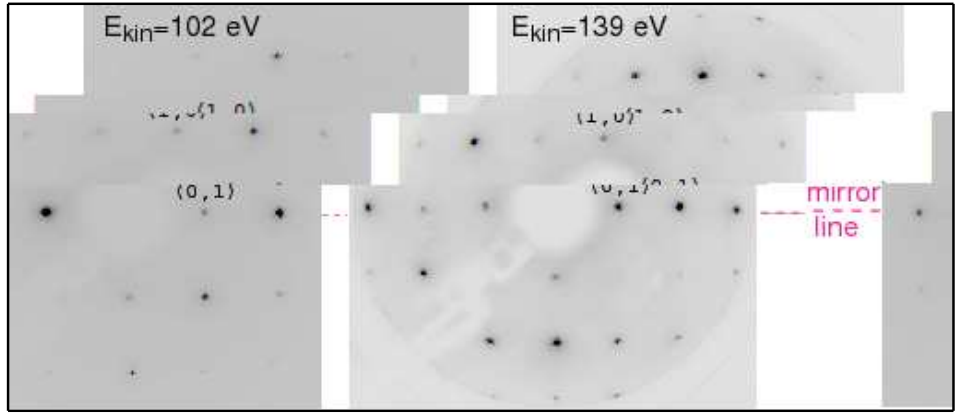}
\caption{(color online) LEED patterns at two different electron beam energies 
for normal incidence on Bi(110) at 110 K. 
The (1,0) and (0,1) diffraction spots are marked.}\label{fig:leedpattern110}
\end{figure}

To obtain intensities of the diffracted beams as a function of electron energy, the following procedure was employed:
A series of images was recorded within the energy range from 30 eV to 300 eV, 
while the energy was increased in steps of 1 eV after every recorded image.
The integrated spot intensity of every single diffracted beam $(h,k)$ was extracted from these images. 
The presence of only one mirror line symmetry for Bi(110) 
leads to technical challenges for the LEED experiment. 
These difficulties are illustrated in Fig.~\ref{fig:leedpattern110} 
that shows two measured LEED patterns taken at different incident energies. 
The pseudo-square pattern of the reciprocal lattice and the missing left/right symmetry are clearly evident.
The up/down symmetry is given by the mirror plane in the crystal 
(the horizontal plane in Fig.~\ref{fig:leedpattern110}). 
It is necessary to align the sample surface perpendicular to the incoming electron beam,
and this is usually done by comparing the IV curves of the symmetry-equivalent beams. 
Here this procedure can only be applied for the up/down angle. 
In order to align the left/right angle we optimized the diffraction spot position on the LEED screen 
until they agreed with the kinematically calculated positions.
We estimate that this approach leads to an error of less than $1^{\circ}$ in the angle of incidence.
In the final data set, the intensities of the symmetry-equivalent beams were averaged.
 
\subsection{Dynamical LEED calculations}\label{sec:4444}

The dynamical LEED intensity calculations were performed using the standard package SATLEED 
(Symmetrized Automated Tensor LEED) by A. Barbieri and M.A. Van Hove\cite{BaNet} 
within the renormalized forward scattering perturbation formalism.
Atomic scattering phase shifts have been calculated using a muffin-tin potential model 
and the standard Barbieri and Van Hove phase shift package.\cite{BaNet}   
The bulk diffraction matrices for the closely spaced bilayers were calculated 
with the combined space method.\cite{VanHove:1979} 
The same muffin-tin radius of 2.87 a.u.\  and phase shifts have been used as in Ref.~\onlinecite{Monig:2005}. 
Phase shifts have been renormalized by the thermal effects of root-mean-square (rms) isotropic vibrational amplitudes. 
Up to 15 ($L=14$) phase shifts have been used because of the strong scattering of the heavy Bi atom ($Z= 83$). 
The muffin-tin constant $V_{0}$ is taken to be energy-independent and is optimized. 
$V_{im}$, the imaginary part of inner potential, also referred to as damping potential or optical potential, 
is taken as 4 eV for the bulk and 4.2 eV for the first 2 overlayers.
The slightly larger value at the surface was chosen to model the presence of dangling bonds,
which increases the electron damping.
The surface potential step of height $V_0$ is located half a long bulk interlayer spacing away from the topmost layer nuclei. The bulk Debye temperature is fixed at 119 K,~\cite{Jezequel:1984} while the Debye temperatures for the first 2 layers are optimized.
Mean-square atomic vibrational amplitudes $<u^{2}>_{T}$ at temperature $T$ for the Debye-Waller factor calculation are derived from Debye temperatures $\Theta_{D}$ according to the following equation: \cite{Pendry:1974} 
\begin{equation}
<u^{2}>_{T}=\frac{9\hbar^{2}}{m_{a}k_{B}\Theta_{D}}\left\{\frac{T^{2}}{\Theta_{D}^{2}}\int^{\frac{\Theta_{D}}{T}}_{0}\frac{xdx}{e^{x}-1}+\frac{1}{4}\right\},
\label{eq:deby}
\end{equation}
where $m_{a}$ is the atomic mass, $\hbar$ Planck's constant and
$k_{B}$ the Boltzmann constant.

In the LEED intensity analysis, agreement between experimental and calculated LEED intensities
is quantified by the widely used Pendry $R$ factor, $R_{P}$, 
which is particularly sensitive to relative peak position and the existence of small peaks.\cite{Pendry:1980} 
The uncertainties in the optimized structural parameters are estimated from the variation around the minimum 
$R_{P \min}$,

\begin{equation}\label{eq:deltaR}
\Delta{R_{P}}=R_{P \min}\times\sqrt{8|V_{im}|/\Delta{E}},
\end{equation}
where $\Delta{E}$ is the total energy range compared in the IV analysis.\cite{Pendry:1980}

\subsection{ {\it Ab initio} calculations}

We have also performed {\it ab initio} calculations of the surface crystal structure of Bi(110).  
The full-potential linearized augmented plane wave method in film-geometry\cite{Krakauer:79.1,Wimmer:81.1} 
as implemented in the FLEUR-code was used and the local density approximation\cite{Moruzzi:78.1}
to the density functional theory was employed. 
Spin-orbit coupling was included in the self-consistent calculations.\cite{Li:90.1}
The evaluation of the surface relaxation has been carried out for the symmetric 14-layer film, 
both, with the inclusion of the spin-orbit coupling (SOC) term and without this term.
Force calculations have been performed for the first four layers without spin-orbit coupling 
while relaxations have been carried out only for the first two interlayer spacings with the inclusion of SOC. 
In the latter evaluations we kept the interlayer spacings
$d_{34}$ and $d_{45}$ equal to those obtained from the force calculation without SOC.
The geometry was chosen such, that both sides of the film were terminated with an intact bilayer.
A wavefunction cutoff of 3.8 a.u.$^{-1}$ was chosen 
and the Brillouin zone was sampled with 32 {\bf k}-points.

\section{RESULTS AND DISCUSSIONS}\label{sec:5555}

\subsection{LEED structure determination}

The LEED pattern of Bi(110) has previously been discussed by Jona.\cite{Jona:1967} 
He defined a pseudo-cubic bulk unit cell and concluded that the unit cell (and hence the LEED pattern) 
should not be exactly rectangular but that the lattice vectors should include an angle 
slightly different from 90$^{\circ}$. 
Our study does not confirm this conclusion.  
Our LEED patterns as presented in Fig.~\ref{fig:leedpattern110} show an exact rectangular net
from careful measurements of the diffraction spots positions and, indeed, 
such an exact rectangle can also be expected from a projection of the bulk reciprocal lattice 
onto the surface.\cite{Hofmann:2006} 
The measured ratio of the two reciprocal unit cell vectors is 0.96(2) 
in good agreement with the expected value of 0.959. 
Moreover, the observed patterns show no indication of any reconstruction of the Bi(110) surface, 
despite the existence of active dangling bond at the surface. 
Apparently, Bi(110) is found to be very different from typical semiconductors surfaces, 
such as Si(100) and Ge(100) which both exhibit 2$\times$1 reconstructions.

The structural and non-structural parameters were optimized for a Bi(110) surface terminated by an intact bilayer.
A termination with a split bilayer was immediately excluded 
due to lack of agreement with the experimental IV curves shown in Figs.~\ref{fig:IV110a} and \ref{fig:IV110b}. 
20 symmetry in-equivalent beams with a total energy range of 3591 eV 
have been analyzed to determine the following structural and non-structural parameters:
the first four interlayer spacings $d_{ij}$ ($j=i+1$; $1 \leq i \leq 4$),
the real part of inner potential $V_{0}$, 
and Debye temperatures $\Theta_{D_{1}}$ and $\Theta_{D_{2}}$ for atoms in the first and second layers, respectively. 
The results of the structural analysis are summarized in Table~\ref{tab:results110}.
Note that the first and the third interlayer spacing
correspond to the small separation (0.21 \AA) between the two layers making up the bilayer in the bulk.  
Their seemingly dramatic relative relaxations are very small in absolute terms.
Also, the forth layer appears to move above the third layer by 0.01 \AA.  
However this very small value is clearly below our detection limit. 
Overall no significant relaxation for the Bi(110) surface is found.
We have tried many possible displacement patterns allowed due to the low surface symmetry. 
However, we found no significant improvement in $R_{P}$ 
when changing the relative distance between the two basis atoms in the first and second layer parallel to the mirror line.
The Debye temperature for the first layer is found to be lower than that of the bulk, 
which is consistent with an early study of Goodman and Somorjai.\cite{Goodman:1970} 
Reduced surface Debye temperatures are a common phenomenon 
reflecting the weaker bonding of surface atoms compared to the bulk.\cite{Walfried:1996} 
The actual numerical values of the surface Debye temperature are an important ingredient for the determination
of the electron-phonon coupling strength from angle-resolved photoemission data.\cite{Ast:2002, Gayone:2005,Kim:2005a} 
Meanwhile the second layer shows a Debye temperature close to the bulk value. 

\begin{table}
\centering
\caption{Optimized parameter values with errors 
for the surface structure of Bi(110) at 110 K and normal incidence. 
Interlayer spacings between the $i$th and $j$th layer are indicated as $d_{ij}$. 
$d_{ij}^{\mathrm{b}}$ is the corresponding interlayer spacing of the truncated bulk at 110 K and
$\Delta d_{ij} = d_{ij}-d_{ij}^{\mathrm{b}}$.
$V_{0}$ is the real part of inner potential. 
$\Theta{_{D_{1}}}$ and $\Theta{_{D_{2}}}$ are the Debye temperatures for the first and the second layer, 
respectively.}\label{tab:results110}

\begin{tabular}{ c c c c }
\\
\hline\hline
Parameters & Starting values & Optimized values & $\Delta d_{ij} / d_{ij}^{\mathrm{b}}$ (\%)\\             
\hline
  $V_{0}$ (eV)    & 8.0     &  $3.5\pm1.50$   & $-$  \\
  $d_{12}$ (\AA) & 0.208 & $0.18\pm0.048$ & $-13\pm23$  \\
  $d_{23}$ (\AA) & 3.064 & $3.06\pm0.043$ & $-0.2\pm1.4$  \\
  $d_{34}$ (\AA) & 0.208 & $0.01\pm0.040$ & $-105\pm19$  \\
  $d_{45}$ (\AA) & 3.064 & $3.20\pm0.046$ & $+4.3\pm1.5$  \\
  $\Theta_{D_{1}}$ (K) & 119 & $95^{+60}_{-25}$ & $-$   \\
  $\Theta_{D_{2}}$ (K) & 119 & $116^{+80}_{-40}$ & $-$  \\
  $R_{P}$  & $-$ & $0.455$ & $-$  \\
\hline\hline
\end{tabular}
\end{table}

The LEED-IV analysis gives a relatively high $R_{P}$ factor of about 0.455 
compared to typical values of 0.1 to 0.3 for clean unreconstructed metal surfaces.
We believe it is due to the structural complexity and low symmetry of the Bi(110) surface
and it is not caused by deviations from normal incidence during the IV measurement.
We simulated non-normal incidence conditions extensively in the LEED-IV calculations
and found that an increase in the incident angle gave a dramatic rise in the $R$ factor
from its minimum at zero or normal incidence.
This shows that the sample is properly aligned.
As seen on open semiconductor surfaces the presence of dangling bonds and
the presence of voids in the open surface structure is a real challenge for the muffin-tin approximation
of the crystal potential 
and could also contribute to the relatively high $R_{P}$ for this surface.
However, the low surface symmetry of Bi(110) gives rise to the large number of non-equivalent beams.
The good agreement between this large experimental data set and the calculated intensity-energy curves, 
as shown in Figs.~\ref{fig:IV110a} and \ref{fig:IV110b}, give us great confidence in the reliability of our results.

\begin{figure}
   \includegraphics[width=4.50in]{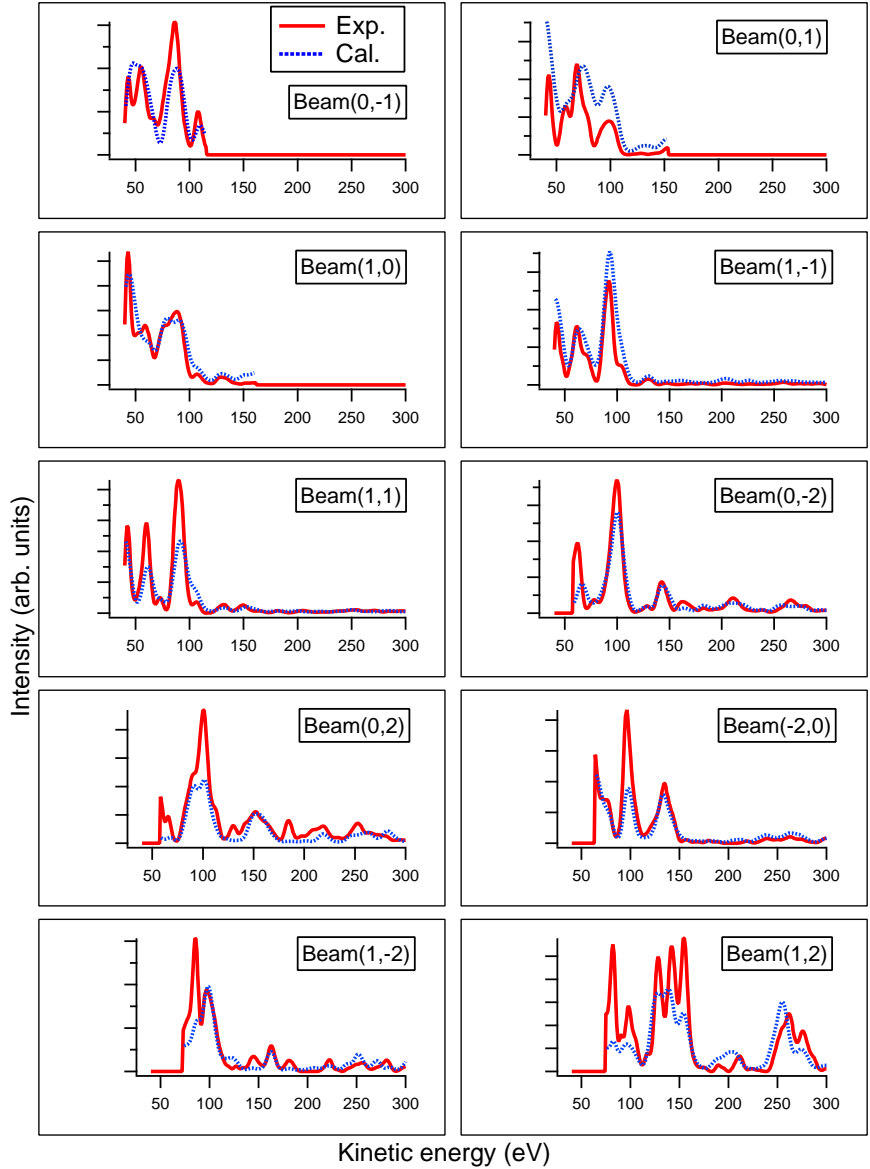}
  \caption{(color online) Comparison of 20 experimental and calculated IV curves for normal incidence on Bi(110) at 110 K. 
  Solid lines show experimental data and dotted lines show calculated data. 
  To be continued in Fig.~\ref{fig:IV110b}.}\label{fig:IV110a}
\end{figure}

\begin{figure}
   \includegraphics[width=4.50in]{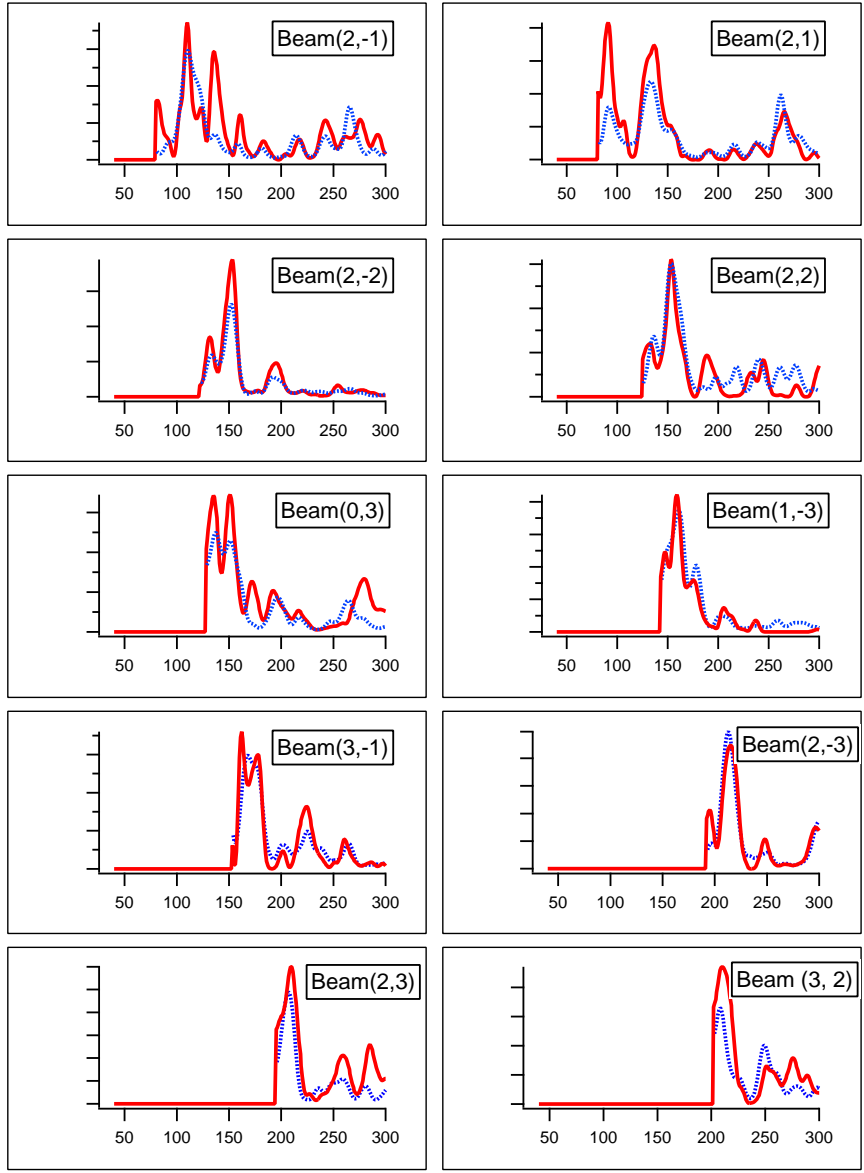}
  \caption{Continuation of Fig.~\ref{fig:IV110a}.}\label{fig:IV110b}
\end{figure}

\begin{figure}
    \centering
        \includegraphics[width=3.50in]{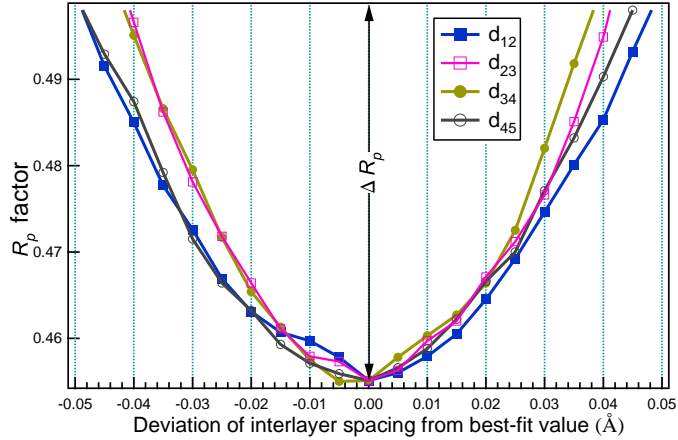}
     \caption{(color online) Error bar determination for the first 4 interlayer spacings based on $\Delta{R_{P}}$ = 0.043 
     and Eq.~\ref{eq:deltaR}.}
    \label{fig:error110}
\end{figure}

The error bars of the optimized parameters were analyzed based on the variation the $R$ factor 
around $R_{P \min}$, $\Delta{R_{P}}$ = 0.043 according to Eq.~\ref{eq:deltaR}.  
The dependence of $R_{P}$ on a change of the interlayer spacings away from their optimized values
is shown in the Fig.~\ref{fig:error110}. 
In this analysis, all other parameters were fixed at their optimized values. 
We can see that all the sensitivity curves take on a parabolic shape. 
The errors for the individual parameters are also listed in the Table~\ref{tab:results110}.

\subsection{Comparison to first-principles calculations}

The first-principle calculations performed for bulk Bi without the inclusion of spin-orbit interaction 
give bulk short and long interlayer spacings of 0.142 \AA\ and 3.087 \AA, respectively. 
Evaluations that include the SOC term lead to a very slight modification of approximately 0.01 \AA\ of these results. 
Our scalar relativistic force calculations give the following values for the first 4 interlayer spacing relaxations at 0 K:
$\Delta d_{12}/d_{12}^{\mathrm{b}} = -62$\%,
$\Delta d_{23}/d_{23}^{\mathrm{b}} = +0.3$\%,
$\Delta d_{34}/d_{34}^{\mathrm{b}} = -105$\%, and
$\Delta d_{45}/d_{45}^{\mathrm{b}} = +4.4$\%.
These results agree reasonably well with those obtained by the LEED-IV analysis at 110 K (see Table~\ref{tab:results110}) 
considering the fact that the absolute distance difference 
between the experimental and calculated first interlayer relaxations of $-13$\% and $-62$\% is only 0.06 \AA.
Both the experiment and theory lead to the contraction of the first interlayer spacing.
For the second interlayer spacing the theory gives a small expansion while the experiment 
shows a small contraction of the spacing. 
However the theoretical result is within the experimental error bar. 
The absolute distance difference between the experimental and calculated second interlayer relaxations 
of 0.015 \AA\ is even smaller than that for the first interlayer spacing.
For the third and fourth interlayer spacings the theory and experiment are in excellent agreement.
The first-principle calculations that include the spin-orbit interaction term lead to 
$\Delta d_{12}/d_{12}^{\mathrm{b}} = -43$\% and
$\Delta d_{23}/d_{23}^{\mathrm{b}} = +0.4$\% for the first and second interlayer spacings respectively. 
These values have been obtained by keeping the interlayer spacings
$\Delta d_{34}/d_{34}^{\mathrm{b}}$ and $\Delta d_{45}/d_{45}^{\mathrm{b}}$ 
equal to those found in the scalar relativistic calculations. 
This shows that the influence of  spin-orbit interaction on the relaxation is small and
probably will not change the values of $d_{34}$ and $d_{45}$ significantly.
Notice, that in the relaxed geometry a change of $6$\% or $0.01$ \AA\ in $\Delta d_{12}/d_{12}$ corresponds to
an energy change of only 0.5  meV per surface atom which is certainly at the limit of our accuracy.

In our force calculations, we also optimized the position of the surface atoms in a plane
parallel to the surface. By symmetry, this movement is then confined to the mirror plane
shown in Fig.~\ref{fig:stru} (a). We notice, that these relaxations are small and do not
exceed $1.0$\% in the top four layers, consistent with the experimental findings.

%

\section{Conclusions}\label{sec:6666}

Our results give a consistent picture of the very low-symmetry surface
geometric structure of Bi(110) by LEED intensity analysis and first-principles calculations.  
Good agreement is reached between experimental LEED and theoretical IV curves. 
No structural reconstruction occurs despite of dangling bonds present at the surface.
No significant absolute value of relaxation is found for the first 4 interlayer spacings. 
The reduced top-layer Debye temperature suggests essentially larger vibrational atomic amplitudes at the surface. 
Experimentally, the approach of sample alignment by calculating the diffraction 
spot positions on the LEED screen is very efficient and can be used for surfaces with low 
symmetry as well as for {\it in-situ} cleaved surfaces.

\begin{acknowledgments}
This work was supported by the American National Science Foundation DMR-0134933,
the Danish National Science Foundation, the Basque Country Government, and by the University of the Basque Country.
\end{acknowledgments}

\bibliography{Bi110StructRefs}

\end{document}